\documentclass{article}
\usepackage{amsmath,amssymb,amsfonts}
\usepackage{authblk}
\usepackage{geometry}
\usepackage{graphicx}
\usepackage{url} 

\usepackage{xcolor}
 
\title{Iterative Ricci-Foster Curvature Flow with GMM-Based Edge Pruning: A Novel Approach to Community Detection}
\author[1,2]{Arsenii Onuchin}
\author[3]{Konstantin Sorokin\thanks{Corresponding author \texttt{ksorokin@hse.ru}}}
\author[4]{Maxim Beketov}
\author[5]{Liubov Tupikina}
\affil[1]{Skolkovo Institute of Science and Technology, Moscow, Russia}
\affil[2]{Laboratory of Complex Networks, Center for Neurophysics and Neuromorphic Technologies, Moscow, Russia}
\affil[3]{International Laboratory of Game Theory and Decision Making, HSE University,
St.\,Petersburg, Russia}
\affil[4]{International Laboratory of Stochastic Algorithms and High-Dimensional Inference, HSE University,
Moscow, Russia}
\affil[5]{Phystech School of Applied Mathematics and Computer Science, Moscow Institute of Physics and Technology, Dolgoprudny, Moscow region, Russia}

\date{}

\begin{document}

\maketitle

\begin{abstract}
Community detection in complex networks is a fundamental problem, open to new approaches in various scientific settings. We introduce a novel community detection method, based on Ricci flow on graphs. Our technique iteratively updates edge weights (their metric lengths) according to their (combinatorial) Foster version of Ricci curvature computed from effective resistance distance between the nodes. The latter computation is known to be done by pseudo-inverting the graph Laplacian matrix. At that, our approach is alternative to one based on Ollivier-Ricci geometric flow for community detection on graphs, significantly outperforming it in terms of computation time. In our proposed method, iterations of Foster-Ricci flow that highlight network regions of different curvature -- are followed by a Gaussian Mixture Model (GMM) separation heuristic. That allows to classify edges into ``strong'' (intra-community) and ``weak'' (inter-community) groups, followed by a systematic pruning of the former to isolate communities. We benchmark our algorithm on synthetic networks generated from the Stochastic Block Model (SBM), evaluating performance with the Adjusted Rand Index (ARI). Our results demonstrate that proposed framework robustly recovers the planted community structure of SBM-s, establishing Ricci-Foster Flow with GMM-clustering as a principled and computationally effective new tool for network analysis, tested against alternative Ricci–Ollivier flow coupled with spectral clustering.
\end{abstract}
\begin{figure}
    \centering
    \includegraphics[width=0.9\linewidth]{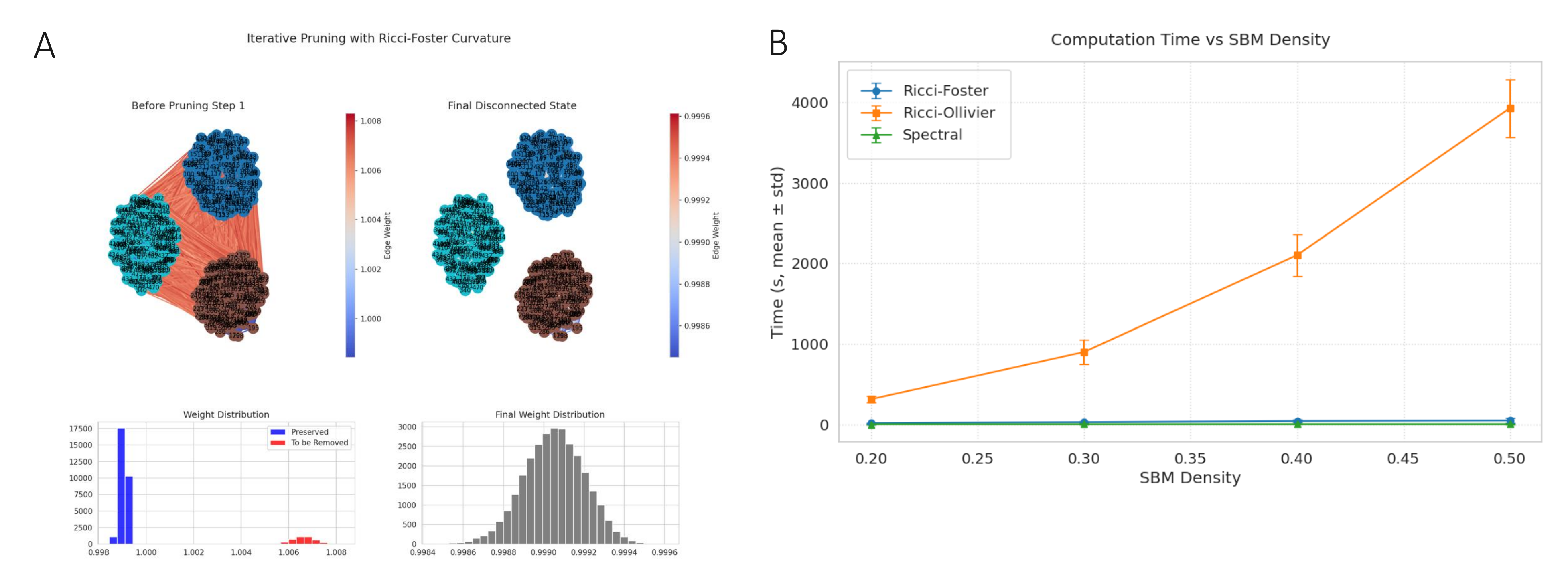}
    \caption{A. Top: Graphs shown before and after pruning edges with the highest weights. Bottom: Distributions of edge weights following iterative Ricci–Foster flow. A Gaussian Mixture Model (GMM) separates the weights into two distinct distributions; all edges corresponding to the rightmost peak are then removed to produce the upper right graph (t-test, p-value$<4.62e-62$) B. Algorithmic performance comparison: Evaluation of the computational efficiency of three clustering approaches—Ricci–Ollivier flow–based clustering, Ricci–Foster flow–based clustering, and spectral clustering. }
    \label{fig:placeholder}
\end{figure}
\section{Introduction}
Community detection is a central problem in the study of complex networks, providing crucial insight into modular organization across domains such as biology, neuroscience, and social systems \cite{fortunato2010community}. Identifying communities—subsets of nodes with dense internal connectivity and sparse external links—enables the characterization of functional organization and structural properties of networks.

In this work, we present a novel community detection algorithm for weighted networks based on the first implementation of Ricci-Foster Flow, which we employ as a clustering method. Ricci-Foster curvature, recently developed for graphs, is constructed from effective resistance distances using the Moore-Penrose pseudoinverse of the Laplacian matrix \cite{dawkins2024riccifoster, devriendt2022discrete}. Our approach iteratively updates edge weights via Ricci-Foster curvature flow, adaptively refining the network's modular structure and accentuating community boundaries. Previous works have also explored Ricci flow for community detection, primarily using the Ollivier curvature definition \cite{ni2019comdet}.

After curvature-based weight evolution, we use the Gaussian Mixture Model (GMM) \cite{bishop2006prml} to statistically separate edges into strong (intra-community) and weak (inter-community) classes, systematically pruning the latter to reveal discrete communities. This geometric-statistical pipeline robustly isolates densely connected modules while reducing spurious inter-cluster links. For benchmarking, we generate synthetic network datasets with planted community structure using the Stochastic Block Model (SBM) \cite{abbe2017sbm, lee2019review} and compare our Ricci-Foster Flow method to spectral clustering \cite{luxburg2007spectral}.

\section{Methods}
\subsection{Ricci-Foster Curvature}
Ricci-Foster curvature offers a spectral-resistance-based definition for edge curvature using the graph Laplacian and its pseudoinverse \cite{dawkins2024riccifoster, devriendt2022discrete}. For each node \(i\), its weighted degree is computed as \(d_i\). Let \(L^+\) denote the Moore-Penrose pseudoinverse of the graph's combinatorial Laplacian matrix \(L\). For an edge \((i,j)\) in the graph, the resistance distance is defined as:
\[
R_{ij} = L^+_{ii} + L^+_{jj} - 2 L^+_{ij}
\]
The Ricci-Foster curvature for the edge \((i,j)\) is then given by:
\[
K_{ij} = \frac{1}{d_i} + \frac{1}{d_j} - \frac{R_{ij}}{w_{ij}}
\]
This value is clipped to the interval \([-1, 1]\) to ensure numerical stability during the flow process \cite{dawkins2024riccifoster}.

\subsection{Ricci Flow on Weighted Graphs}
Ricci flow, first introduced for manifolds \cite{Hamilton1982}, is a process that evolves the network's geometry by updating edge weights based on their curvature \cite{bai2020ollivier, karampour2025discrete}. Given the Ricci-Foster curvature \(\kappa_{uv}^{(t)}\) for each edge \((u,v)\) at iteration \(t\), the edge weights are updated using the following rule \cite{dawkins2024riccifoster}:
\[
w_{uv}^{(t+1)} = \max\left(\epsilon, w_{uv}^{(t)} \cdot \big( 1 - \eta \kappa_{uv}^{(t)} \big) \right)
\]
Here, \(\eta > 0\) is the learning rate, and \(\epsilon > 0\) is a minimum weight threshold to prevent edge weights from becoming zero. After each iteration, the edge weights are normalized to preserve the total weight sum of the graph:
\[
w_{uv}^{(t+1)} \leftarrow \frac{|E|}{\sum_{(u',v') \in E} w_{u'v'}^{(t+1)}} \cdot w_{uv}^{(t+1)}.
\]
This iterative process progressively modifies the graph structure, sharpening the boundaries between modules and making community detection more effective.

\subsection{Algorithm Description}

The proposed algorithm first applies Ricci-Foster Flow to update edge weights, which geometrically reshapes the network to emphasize its modular structure. Following this, an iterative pruning procedure based on a Gaussian Mixture Model is used to systematically remove inter-community edges until distinct clusters are isolated.

\begin{enumerate}
    \item \textbf{Ricci-Foster Flow Stabilization}: Apply the Ricci-Foster Flow procedure to the graph for a fixed number of iterations. This results in updated edge weights that encode geometric information about the graph structure.

    \item \textbf{Iterative Pruning Cycles}: Repeat the pruning process for a set number of cycles or until the graph becomes disconnected.
    \begin{itemize}
        \item \textbf{Edge Weight Extraction}: Collect the current edge weights into a vector.
        \item \textbf{Gaussian Mixture Clustering}: Fit a two-component Gaussian Mixture Model (GMM) to the edge weights. This model probabilistically classifies each edge into one of two groups based on its weight. The GMM is represented as \(p(w_e) = \sum_{k=1}^2 \pi_k\, \mathcal{N}\big(w_e \mid \mu_k, \sigma_k^2\big)\), where \(\pi_k\) are mixture weights, \(\mu_k\) are means, and \(\sigma_k^2\) are variances \cite{bishop2006prml}.
        \item \textbf{Component Assignment and Pruning}: Assign each edge to its most likely GMM component. The component with the lower mean is identified as representing weaker, inter-community edges. All edges belonging to this component are removed from the graph.
        \item \textbf{Connectivity Check}: If the pruned graph is disconnected, its connected components are declared as the final communities, and the process terminates. Otherwise, Ricci-Foster flow is reapplied to the updated graph for another cycle of pruning.
    \end{itemize}
    \item \textbf{Stochastic Block Model (SBM)}: To test the algorithm, we use the Stochastic Block Model (SBM) to generate random graphs with a known, planted community structure \cite{abbe2017sbm, lee2019review}. In an SBM graph, nodes are partitioned into disjoint clusters, and edges are created with a higher probability \(p_{in}\) for nodes within the same community and a lower probability \(p_{out}\) for nodes in different communities. All edges are initially assigned a weight of $1$.
\end{enumerate}

\section{Results}

The proposed network clustering algorithm was evaluated on synthetic networks generated using the Stochastic Block Model (SBM) with \(n=60\) nodes partitioned into \(k=3\) communities. The graph parameters were set to an intra-cluster edge probability of SBM graph \(p_{in}=0.7\) and an inter-cluster probability of \(p_{out} = 0.07\). The results demonstrate that the algorithm reliably detects and recovers the underlying community structure planted in these networks.

\section{Discussions and Conclusions}
One of the principal contributions of this study is the development of a novel statistical method for partitioning a network into community structures. In comparison with existing approaches, such as the Ollivier–Ricci (OR) method, the proposed algorithm exhibits substantially greater scalability: increases in the number of edges have minimal impact on computational performance (Figure 1B).

While the Ricci–Foster flow offers significant advantages as a framework for community detection, certain limitations emerge in graphs with relatively homogeneous edge weight distributions. As part of future research, we aim to enhance the robustness of the algorithm by conducting extensive evaluations on diverse real-world network datasets.

\section*{Code Availability}
The Python code used to implement the Ricci-Foster Flow algorithm and reproduce the results in this paper is publicly available on GitHub at \url{https://github.com/bekemax/Ricci-flow-resistance}.

\section{Acknowledgements}

This work was done during Summer Research Programme at MIPT -- LIPS-25 (Phystech School of applied mathematics and computer science). The work of M. Beketov and K. Sorokin was prepared within the framework of the HSE University Basic Research Program. This research was supported in part with computational resources of HPC facilities at HSE University \cite{kostenetskiy2021hpc}. The work of L. Tupikina was supported by the ``Priority 2030'' strategic academic leadership program.

\end{document}